\documentclass[12pt]{article}

\usepackage{amsmath,amssymb}%,mathrsfs}
\newcommand{\mathscr}{\cal}

\begin{document}

% End of proof symbol
\renewcommand{\square}{\vrule height 1.5ex width 1.2ex depth -.1ex }

% Define symbols for identity, complex, real and natural numbers.

% Identity using bbold12:
%\font\bbold=bbold12
%\newcommand{\II}{{\hbox{\bbold 1}}}
% Alternatively, macro for identity lifted from REVTeX 3.0:
\newcommand{\II}{\leavevmode\hbox{\rm{\small1\kern-3.8pt\normalsize1}}}

\newcommand{\CC}{{\mathbb C}}
\newcommand{\RR}{{\mathbb R}}
\newcommand{\NN}{{\mathbb N}}
\newcommand{\QQ}{{\mathbb Q}}
\newcommand{\ZZ}{{\mathbb Z}}

% Various classes of smooth compactly supported functions            

\newcommand{\CoinfM}{C_0^\infty(M)}
\newcommand{\CoinfN}{C_0^\infty(N)}
\newcommand{\Coinfd}{C_0^\infty(\RR^d\backslash\{ 0\})}
\newcommand{\Coinf}[1]{C_0^\infty(\RR^{#1}\backslash\{ 0\})}
\newcommand{\CoinX}[1]{C_0^\infty({#1})}
\newcommand{\Coin}{C_0^\infty(0,\infty)}

% Various Theorem-like environments

\newtheorem{Thm}{Theorem}[section]
\newtheorem{Def}[Thm]{Definition}
\newtheorem{Lem}[Thm]{Lemma}
\newtheorem{Prop}[Thm]{Proposition}
\newtheorem{Cor}[Thm]{Corollary}

% Label equations by section and number-within-section

\renewcommand{\theequation}{\thesection.\arabic{equation}}
\newcommand{\sect}[1]{\section{#1}\setcounter{equation}{0}}

% Calligraphic characters (N.B. to use rsfs use \mathscr) 

\newcommand{\DD}{{\mathscr D}}
\newcommand{\EE}{{\mathscr E}}

\newcommand{\OO}{{\cal O}}

% Boldface characters

\newcommand{\etb}{{\boldsymbol{\eta}}}

% Miscellaneous

\newcommand{\Dal}{\fbox{\phantom{${\scriptstyle *}$}}}

\newcommand{\Ran}{{\rm Ran}\,}
\newcommand{\supp}{{\rm supp}\,}

\newcommand{\stack}[2]{\substack{#1 \\ #2}}

\begin{titlepage}
\renewcommand{\thefootnote}{\fnsymbol{footnote}}

%\rightline{BUTP-95/31}
\vspace{0.1in}
\LARGE
\center{A Unique Continuation Result for Klein--Gordon Bisolutions
on a 2-dimensional Cylinder} 
\Large
%\vspace{0.2in}

\vspace{0.2in}
\center{C.J. Fewster\footnote{E-mail: {\tt cjf3@york.ac.uk}}}
\center{Department of Mathematics, University of York, \\
Heslington, York YO1 5DD, United Kingdom.}

\normalsize
\center{April 3, 1998}

\begin{abstract}
We prove a novel unique continuation result for weak bisolutions to the
massive Klein--Gordon equation on a 2-dimensional cylinder $M$. Namely,
if such a bisolution vanishes in a neighbourhood of a `sufficiently
large' portion of a 2-dimensional surface lying parallel to the diagonal
in $M\times M$, then it is (globally) translationally invariant. The
proof makes use of methods drawn from Beurling's theory of
interpolation. An application of our result to quantum field theory on
2-dimensional cylinder spacetimes will appear elsewhere.
\end{abstract}

\end{titlepage}

\sect{Introduction}

The term `unique continuation' covers a variety of results concerning
the support properties of solutions to partial differential equations
(or inequalities); see, for example Chapters~8, 17 and~28 
of~\cite{Horm}. In this short
paper we will study an apparently new variation on this theme concerning  
weak bisolutions to the massive Klein--Gordon equation on
a 2-dimensional cylinder $M$. We will show that if such a bisolution
vanishes in a neighbourhood of a sufficiently large portion of a
2-dimensional surface lying parallel to the diagonal in
$M\times M$ then it must be translationally invariant. This result was
motivated by a study of quantum field theory on cylinder
spacetimes~\cite{F,FHK} in which it is used to deduce global
information from knowledge of the expectation value of the commutator   
$[\widehat{\phi}(x),\widehat{\phi}(x')]$ of a quantum field
$\widehat{\phi}$ for nearby points $x$ and $x'$. 

To formulate our result precisely, define $M$ to be the cylinder formed
as the quotient of $\RR^2$ by the translation $z\mapsto z+2\pi$, where
$(t,z)$ are Cartesian coordinates on $\RR^2$, and let $P$ be the
Klein--Gordon operator on $M$ given by 
\begin{equation}
P = \frac{\partial^2}{\partial t^2} - \frac{\partial^2}{\partial z^2}+\mu
\label{eq:KG}
\end{equation}
in local $(t,z)$-coordinates, where $\mu$ is a real constant. 
By a weak $P$-bisolution, we mean a bidistribution 
$\Gamma\in \DD'(M\times M)$ such that 
\begin{equation}
\Gamma(Pf\otimes g) = \Gamma(f\otimes Pg) = 0
\end{equation}
for all test functions $f,g\in\DD(M)$. Such bisolutions play a well
known and important r\^ole in linear scalar quantum field theory as
`two-point functions'~\cite{Wald}. We shall prove that weak $P$-bisolutions
exhibit a unique continuation property modulo translational invariance
across 2-dimensional surfaces in $M\times M$ of form 
\begin{equation}
\label{eq:surf}
S_{pp'} =\left\{
\left(T_{(\tau,\zeta)}p,T_{(\tau,\zeta)}p'\right)\in M\times M
\mid \tau\in[0,\pi],\quad\zeta\in[0,2\pi]\right\}
\end{equation}
for some fixed $p,p'\in M$, where $T_{(\tau,\zeta)}$ translates points by
\begin{equation}
T_{(\tau,\zeta)} q(t,z) = q(t+\tau,z+\zeta)
\end{equation}
and $q:\RR^2\to M$ is the defining
quotient map. These compact surfaces lie within
(2-dimensional) planes parallel to the diagonal in the product manifold 
$M\times M$. Our result is the following. 
\begin{Thm} \label{Thm:main}
Let $\mu$ be any real number other than a negative integer
or zero and define $P$ by Eq.~(\ref{eq:KG}). If $\Gamma$ is a weak
$P$-bisolution vanishing on an open neighbourhood of some surface $S_{pp'}$ of
the form~(\ref{eq:surf}) then
$\Gamma$ is translationally invariant. 
\end{Thm}
Here, $\Gamma$ is said to be translationally invariant if
$\Gamma(T_{(\tau,\zeta)}f\otimes T_{(\tau,\zeta)} g)=\Gamma(f\otimes g)$ 
for all test functions $f,g\in\DD(M)$ and all $(\tau,\zeta)\in\RR^2$,
where the translation $T_{(\tau,\zeta)}$ acts on test functions by
\begin{equation}
\left( (T_{(\tau,\zeta)}f)\circ q\right)(t,z) =
\left( f\circ q\right)(t-\tau,z-\zeta).
\end{equation}

Theorem~\ref{Thm:main} could also be stated by saying that the
restriction of a weak $P$-bisolution $\Gamma$ to any neighbourhood of
$S_{pp'}$ determines $\Gamma$ uniquely up to the addition of a
translationally invariant bisolution. We remark that nontrivial
translationally invariant bisolutions which vanish near $S_{pp'}$ are
easily constructed from any weak solution $\varphi$ vanishing near
$q(0,0)$ by writing (rather loosely)
\begin{equation}
\Gamma(t,z;t',z')=\varphi(t-t'+\tau,z-z'+\zeta)
\end{equation}
where $\tau,\zeta$ are chosen so that $p'=T_{(\tau,\zeta)}p$.  
In the hypotheses of Theorem~\ref{Thm:main} the
condition $\mu\not=0$ is essential, while the requirement that $\mu$
should not equal a negative integer is primarily a technical convenience
and can doubtless be removed. To see that the result fails if $\mu=0$,
it suffices to note that if $G\in\DD'(\RR\times\RR)$ is a doubly
$2\pi$-periodic distribution, then (informally)
\begin{equation}
\Gamma(t,z;t',z') = G(t-z,t'-z')
\end{equation}
is a weak $P$-bisolution in this case.  
The requirement that $\Gamma$ vanish in a neighbourhood of some
surface $S_{pp'}$ of the form~(\ref{eq:surf}) restricts $G(u,u')$ only in
the vicinity of some diagonal $u=u'+{\rm const}$, so one may easily
construct nontranslationally invariant bisolutions $\Gamma$ vanishing
near $S_{pp'}$ if $\mu=0$. 

We will prove Theorem~\ref{Thm:main} in Section~\ref{sect:main} using a
result on exponential series established in Section~\ref{sect:Beurling}
which draws on Beurling's theory of interpolation~\cite{Beu}.
(These ideas also underpin Shannon-type theorems in signal processing
(see, e.g.,~\cite{Landau}.) As mentioned above, an application of
Theorem~\ref{Thm:main} to quantum field theory on cylinder spacetimes
will appear elsewhere~\cite{F,FHK}.

\sect{A Result on Exponential Series}\label{sect:Beurling}

In this section, we use ideas and methods drawn from Beurling's theory
of interpolation~\cite{Beu} to obtain a result on exponential series
which will be the key to the continuation result proved in the next
section. 

Let $\Lambda\subset\CC$ be a countable set with at most finitely
many non-real elements. For each complex sequence 
$c=\{c_\lambda\}_{\lambda\in\Lambda}$ labelled by $\Lambda$ and
obeying
\begin{equation}
\sum_{\lambda\in\Lambda} |c_\lambda| < \infty,
\end{equation}
i.e., $c\in\ell^1(\Lambda)$,
there is a corresponding exponential series
\begin{equation}
\omega_{c}(x) = \sum_{\lambda\in\Lambda} c_\lambda e^{i\lambda x}
\end{equation}
converging uniformly on compact subsets of $\RR$ and
a distribution $\omega_c\in\DD'(\RR)$ defined by 
\begin{equation}
\omega_c(f) = \int_{-\infty}^\infty \! dx\, \omega_c(x) f(x)
=\sum_{\lambda\in\Lambda} c_\lambda \widehat{f}(\lambda)
\end{equation}
for $f\in\DD(\RR)$. Here, the Fourier transform of $f$ is defined by
\begin{equation}
\widehat{f}(\lambda) = \int_{-\infty}^\infty \! dx\, 
e^{i\lambda x} f(x).
\end{equation}
It is essentially straightforward to show that the map
\begin{eqnarray}
\omega:\ell^1(\Lambda) &\longrightarrow& \DD'(\RR) \nonumber \\
c & \longmapsto & \omega_c
\end{eqnarray}
is an injection, i.e., if $\omega_c(f)=0$ for all $f\in\DD(\RR)$, then
$c_\lambda=0$ for all $\lambda\in\Lambda$. In this section we will
establish a condition for $\omega$ to be injective from $\ell^1(\Lambda)$
to $\DD'(E)$ for $E$ a finite interval in $\RR$. We will require the
following:
\begin{Def} A (countable) set $\Lambda\subset\CC$ is said to be {\em
uniformly discrete} if there exists $\delta>0$ such that 
$|\lambda-\lambda'|>\delta$
for all pairs $\lambda,\lambda'$ of distinct elements of $\Lambda$. 
If $\Lambda$ is uniformly discrete and contains at most finitely
many non-real elements, then we define the {\em upper uniform density}
$D^+(\Lambda)$ of $\Lambda$ to be
\begin{equation}
D^+(\Lambda) = \lim_{r\to\infty} \frac{N(r)}{r}
\end{equation}
where $N(r)$ is the maximum number of elements of $\Lambda$ contained
in any interval of form $[x,x+r]$, $x\in\RR$. 
\end{Def}
We remark that $N(r)/r$ has a limit as $r\to\infty$ because $N(r)$ is
subadditive and $N(r)/r$ is bounded above (by uniform discreteness). 

Our main result in this section is based on the following, which
is a slight modification of a result due to Beurling in~\cite{Beu}.
\begin{Prop} \label{Prop:Beurling}
If $\Lambda\subset\CC$ is uniformly discrete and has at most finitely
many non-real elements and $E$ is any real interval of length greater
than $2\pi D^+(\Lambda)$, then there exist test functions 
$\{f_\lambda\}_{\lambda\in\Lambda}$ in $\DD(E)$ such that
\begin{equation}\label{eq:on}
\widehat{f_\lambda}(\lambda')= 
\left\{\begin{array}{cl} 1 & \lambda=\lambda' \\ 0 & \lambda\not=\lambda'
\end{array}\right.
\end{equation}
for $\lambda,\lambda'\in\Lambda$. 
\end{Prop}
In the case $\Lambda\subset\RR$, this follows
directly from the proof of Theorem~1 in~\cite{Beu}. In that argument, 
Beurling constructs analytic functions
$\widehat{f_\lambda}$ [which he denotes $g_\lambda$]
obeying Eq.~(\ref{eq:on}) and which are of the
correct exponential type to be Fourier transforms of functions supported
in $E$. By construction, these functions are in fact Fourier
transforms of functions in $\DD(E)$, although continuity suffices for
Beurling's purposes. The result is easily  
extended to include finitely many non-real elements using the ideas of
Lemma~4 in~\cite{Beu}. Beurling's result has the following immediate 
consequence:
\begin{Cor} \label{cor:Beurling} 
If $\Lambda$ and $E$ obey the hypotheses of 
Theorem~\ref{Prop:Beurling} then $\omega$ is injective as a map from 
$\ell^1(\Lambda)$ to $\DD'(E)$. 
\end{Cor} 
{\noindent\em Proof:} Applying Proposition~\ref{Prop:Beurling}, we have
$c_\lambda= \omega_c(f_\lambda)$ for all $\lambda\in\Lambda$. $\square$

Our main result here is a slight extension of
Proposition~\ref{Prop:Beurling} 
to address more general countable sets $\Lambda$ which may contain
accumulation points.  

\begin{Thm} \label{thm:Beurlingplus}
Suppose $\Lambda\subset\CC$ is countable with at most
finitely many non-real elements. If there exists $R>0$ such that
$\Lambda_R= \{\lambda\in\Lambda\mid~|\lambda|>R\}$ is uniformly
discrete then $\omega:\ell^1(\Lambda)\to\DD'(E)$ is injective whenever
$E$ is an interval of length greater than $2\pi D^+(\Lambda_R)$.
\end{Thm}
{\noindent\em Proof:} Let $a$ and $b$ be any real numbers such that
$a>b>2\pi D^+(\Lambda_R)$ and suppose a sequence
$c=\{c_\lambda\}_{\lambda\in\Lambda}$ exists such that $\omega_c(f)=0$ 
for all $f\in \DD(0,a)$. We will show that $c_\lambda=0$ for all 
$\lambda\in\Lambda$, thus establishing that $\omega$ is injective
from $\ell^1(\Lambda)$ to $\DD'(0,a)$. Let
\begin{equation}
\varphi(x) = \sum_{\lambda\in\Lambda_R}
c_\lambda e^{i\lambda x}
\qquad{\rm and}\qquad
\psi(x) =\sum_{\lambda\in \Lambda\backslash\Lambda_R} 
c_\lambda e^{i\lambda x}
\end{equation}
so that $\omega_c=\varphi+\psi$. 

For each $\mu\in \Lambda\backslash\Lambda_R$,
we apply Proposition~\ref{Prop:Beurling} to the set $\Lambda_R\cup\{\mu\}$ and
the interval $(0,b)$ to obtain a test function $f_\mu\in\DD(0,b)$ such
that $\widehat{f_\mu}(z)$ vanishes for $z\in\Lambda_R$ and equals unity for
$z=\mu$. Now consider the 1-parameter family of test functions 
$(T_\alpha f_\mu)(x)=f_\mu(x-\alpha)$ in $\DD(0,a)$ for $\alpha\in[0,a-b]$. 
We have $\omega_c(T_\alpha f_\mu)=0$
by hypothesis on $\omega_c$; moreover
\begin{equation}
\varphi(T_\alpha f_\mu) = \sum_{\lambda\in\Lambda_R} c_\lambda 
e^{i\lambda \alpha} \widehat{f_\mu}(\lambda)
=0
\end{equation}
by definition of $f_\mu$. Thus 
\begin{equation}
\psi(T_\alpha f_\mu) = \sum_{\lambda\in\Lambda\backslash\Lambda_R}
 e^{i\lambda\alpha} c_\lambda \widehat{f_\mu}(\lambda),
\label{eq:psa}
\end{equation}
vanishes for all $\alpha\in[0,a-b]$. Since the right-hand side of 
Eq.~(\ref{eq:psa}) extends to an analytic function of $\alpha$ [because 
$\Lambda\backslash\Lambda_R$ is a bounded subset of $\CC$] we 
deduce that it must vanish identically. It follows that $c_\lambda
\widehat{f_\mu}(\lambda)=0$ for all $\lambda\in
\Lambda\backslash\Lambda_R$ 
and in particular that $c_\mu=0$ [because $\widehat{f_\mu}(\mu)=1$]. 

We have now shown that $c_\mu=0$ for all $\mu\in
\Lambda\backslash\Lambda_R$. 
To conclude the proof, we apply Corollary~\ref{cor:Beurling} to the
set $\Lambda_R$ and the interval $(0,a)$ to show that $c_\lambda=0$
for all $\lambda\in\Lambda_R$. Thus $\omega$ is injective as a map
from $\ell^1(\Lambda)$ to $\DD'(0,a)$ and hence from $\ell^1(\Lambda)$ 
to $\DD'(E)$ whenever $E$ is a real 
interval of length greater than $2\pi D^+(\Lambda_R)$. 
$\square$

\section{Proof of Theorem~\ref{Thm:main}}
\label{sect:main}

We begin with some preliminaries. Let $\mu$ be any real number other
than zero or a negative integer, and let $\xi_{n\epsilon}$ (for
$n\in\ZZ$, $\epsilon=\pm$) denote the distribution acting on
$f\in\DD(M)$ by
\begin{equation}
\xi_{n\epsilon}(f) = \int_{\RR\times[0,2\pi]} dt\,dz\,
e^{-i\epsilon\omega_n t+inz} (f\circ q)(t,z)
\end{equation}
where $q$ is the quotient map defining $M$ and $\omega_n$ is given by
\begin{equation}
\omega_n^{1/2} =\left\{\begin{array}{cl} |n^2+\mu|^{1/4} & n^2>-\mu
\\ e^{i\pi/4}|n^2+\mu|^{1/4} & n^2<-\mu. \end{array}\right.
\end{equation}
If $\mu<0$, finitely many of
the $\omega_n$ can be imaginary. Equivalently, 
\begin{equation}
\xi_{n\epsilon}(f) = \widehat{f}(\epsilon\omega_n,n)
\end{equation}
where the Fourier transform $\widehat{f}$ of $f\in\DD(M)$ is
\begin{equation}
\widehat{f}(\omega,n)= \int_{\RR\times[0,2\pi]} dt\,dz\,
e^{-i\omega t+inz} (f\circ q)(t,z).
\end{equation}
Each $\xi_{n\epsilon}$ is a weak $P$-solution for the corresponding
value of $\mu$; moreover, any weak $P$-bisolution $\Gamma$ may be
expanded in terms of the $\xi_{n\epsilon}$:
\begin{equation}
\Gamma = \sum_{\stack{n,n'\in\ZZ}{\epsilon,\epsilon'=\pm}}
\gamma_{nn'}^{\epsilon\epsilon'}
\xi_{n\epsilon}\otimes\xi_{n'\epsilon'},
\label{eq:Gexp}
\end{equation}
where the coefficients $\gamma_{nn'}^{\epsilon\epsilon'}$ grow no
faster than polynomially in $n,n'$, so the sum converges in the weak-$*$
topology on $\DD'(M\times M)$. Clearly,
\begin{equation}
\Gamma(T_{(\tau,\zeta)}f_1\otimes T_{(\tau,\zeta)}f_2) = 
\sum_{\stack{n,n'\in\ZZ}{\epsilon,\epsilon'=\pm}}
\gamma_{nn'}^{\epsilon\epsilon'} 
e^{i(n+n')\zeta-i(\epsilon\omega_n+\epsilon' \omega_{n'})\tau}
\xi_{n\epsilon}(f_1)\xi_{n'\epsilon'}(f_2),
\label{eq:tran}
\end{equation}
from which it follows that $\Gamma$ is translationally invariant if and
only if the only nonzero coefficients $\gamma_{nn'}^{\epsilon\epsilon'}$
are those on the anti-diagonal, with $n=-n'$, $\epsilon=-\epsilon'$. 

We now prove Theorem~\ref{Thm:main}, which states that a weak
$P$-bisolution is translationally invariant if it vanishes on a
neighbourhood of a diagonal set $S_{pp'}$ of the
form~(\ref{eq:surf}) for $p,p'\in M$. 

{\noindent\em Proof of Theorem~\ref{Thm:main}:} We express $\Gamma$ in
the form~(\ref{eq:Gexp}) and show that all coefficients
$\gamma_{nn'}^{\epsilon\epsilon'}$ other than those on the anti-diagonal
must vanish if $\Gamma$ vanishes in a neighbourhood $N$ of a diagonal
set $S_{pp'}$. First note that, by compactness of $S$,
there are open neighbourhoods $\OO_1$ of $q(t_1,z_1)$ and $\OO_2$ of
$q(t_2,z_2)$ in $M$ such that $N$ contains the translates
$T_{(\tau,\zeta)}\OO_1\times T_{(\tau,\zeta)}\OO_2$ for $\tau\in E$,
$\zeta\in [0,2\pi]$ where $E$ is a real interval of length greater than
$\pi$. Thus, fixing $f_i\in\DD(\OO_i)$ we have
\begin{equation}
F(\tau,\zeta) \stackrel{{\rm def}}{=}
\Gamma(T_{(\tau,\zeta)}f_1\otimes T_{(\tau,\zeta)}f_2) = 0
\end{equation}
for all $\tau,\zeta$ in these ranges. If we now take Fourier components
of $F$, defining $F_N(\tau)$ ($N\in\ZZ)$ by
\begin{equation}
F_N(\tau) = \frac{1}{2\pi}\int_0^{2\pi}\!d\zeta\, e^{-iN\zeta} F(\tau,\zeta) ,
\end{equation}
we obtain (comparing with Eq.~(\ref{eq:tran}))
\begin{eqnarray}
F_N(\tau) 
&=&
\sum_{(n,n')\in I_N^+} \left(
e^{-i\alpha(n,n')\tau}p_{nn'}^+(f_1\otimes f_2)
+e^{i\alpha(n,n')\tau}p_{nn'}^-(f_1\otimes f_2) \right)  \nonumber \\
&&
+ \sum_{(n,n')\in I_N} e^{-i\beta(n,n')\tau} q_{nn'}(f_1\otimes f_2).
 \label{eq:GammaN}
\end{eqnarray}
In this expression, the distributions
$p^\pm_{nn'},q_{nn'}\in\DD'(M\times M)$ are defined by
\begin{equation}
p^\epsilon_{nn'} = \left\{
\begin{array}{cl} \gamma_{nn'}^{\epsilon\epsilon}\xi_{n\epsilon}\otimes
\xi_{n'\epsilon} + \gamma_{n'n}^{\epsilon\epsilon}\xi_{n'\epsilon}
\otimes\xi_{n\epsilon} & n\not= n' \\
\gamma_{nn}^{\epsilon\epsilon}\xi_{n\epsilon}\otimes\xi_{n\epsilon} & n=n'
\end{array}\right.
\end{equation}
and
\begin{equation}
q_{nn'} = \gamma^{+-}_{nn'} \xi_{n+}\otimes\xi_{n'-}
+ \gamma^{-+}_{n'n}\xi_{n'-}\otimes\xi_{n+},
\end{equation}
the frequencies $\alpha(n,n')$ and $\beta(n,n')$ are given by
\begin{eqnarray}
\alpha(n,n') &=& \omega_n+\omega_{n'}  \\
\beta(n,n') &=& \omega_n-\omega_{n'} 
\end{eqnarray}
and the summations are performed over the sets 
$I_N=\{(n,n')\in\ZZ\times\ZZ\mid n+n'=N\}$ and 
$I_N^+=\{(n,n')\in I_N\mid n\ge n'\}$.

{}From Eq.~(\ref{eq:GammaN}) we see that each $F_N(\tau)$ is an
absolutely convergent exponential series with frequencies
\begin{equation}
\Lambda^{(N)} = \{\pm\alpha(n,n')\mid (n,n')\in I_N^+\}\cup
\{\beta(n,n')\mid (n,n')\in I_N\},
\end{equation}
of which at most finitely many are non-real. Furthermore, each 
$F_N(\tau)$ vanishes on a real interval of length greater than $\pi$. We 
will apply the results of Section~\ref{sect:Beurling}, using the
$\pm\alpha(n,n')$ as the uniformly discrete part of the spectrum. The
$\beta(n,n')$ in $\Lambda^{(N)}$ belong to a compact set 
(which depends on $N$) and exhibit accumulation points.  

Starting with $N\not=0$, we note that there is exactly one term
in the series~(\ref{eq:GammaN}) for each frequency in $\Lambda^{(N)}$.
There are two cases. 

{\noindent\em Case I: $\mu>0$.} In this case $\Lambda^{(N)}\subset\RR$.
Owing to the inequalities
\begin{equation}
\alpha(n,n')\ge \sqrt{(n+n')^2+4\mu}>|n+n'|
\ge |\beta(n,n')| \ge 0
\end{equation}
and the fact that $\alpha(n,N-n)=2|n|+O(1)$ as $|n|\to\infty$, the
set 
\begin{equation}
\Lambda^{(N)}_{|N|}=\{\lambda\in\Lambda^{(N)}\mid |\lambda|>|N|\}
\end{equation} 
is uniformly discrete with uniform upper density
$D^+(\Lambda^{(N)}_{|N|})=\frac{1}{2}$. Theorem~\ref{thm:Beurlingplus}
immediately entails that the restriction of $F_N(\tau)$ to
any real interval of length greater than $\pi$ uniquely determines
the coefficients in the series. Accordingly, 
$p^\pm_{nn'}(f_1\otimes f_2)=0$ for $(n,n')\in I_N^+$ and
$q_{nn'}(f_1\otimes f_2)=0$ for $(n,n')\in I_N$.

{\noindent\em Case II: $\mu<0$ with $-\mu\not\in\NN$.} 
In this case $\Lambda^{(N)}$ contains finitely
many non-real frequencies arising when $\min\{|n|,|n'|\}<|\mu|^{1/2}$. 
For $|n-n'|>|n+n'|+2|\mu|^{1/2}$ the inequalities
\begin{equation}
\alpha(n,n')>\sqrt{(n+n')^2-2|n+n'||\mu|^{1/2}}>|\beta(n,n')|>|n+n'|
\end{equation}
hold and since $\alpha(n,N-n)=2|n|+O(1)$ as $|n|\to\infty$ the
set 
\begin{equation}
\Lambda^{(N)}_{R}=\{\lambda\in\Lambda^{(N)}\mid |\lambda|>R\}
\end{equation} 
is uniformly discrete with $D^+(\Lambda^{(N)}_R)=\frac{1}{2}$, where
$R=\sqrt{N^2-2|N||\mu|^{1/2}}$.  
As in Case~I, Theorem~\ref{thm:Beurlingplus} entails that 
$p^\pm_{nn'}(f_1\otimes f_2)=0$ for $(n,n')\in I_N^+$ and
$q_{nn'}(f_1\otimes f_2)=0$ for $(n,n')\in I_N$.

If $N=0$, the above analysis is modified slightly because there are
infinitely many zero frequency terms (since $\beta(n,n')=0$ for all
$(n,n')\in I_0$). Nonetheless, all nonzero frequency terms must vanish 
by identical arguments to those above, so 
$p^\pm_{nn'}(f_1\otimes f_2)=0$ for $(n,n')\in I_0^+$. The
total zero frequency coefficient also vanishes, so
\begin{equation}
\sum_{n\in\ZZ} q_{n\,-n}(f_1\otimes f_2) = 0.
\label{eq:zeroco}
\end{equation}

To summarise, letting $N$ run through $\ZZ$, 
we have $p_{nn'}^\pm(f_1\otimes f_2)=0$ for all $n\ge n'$, 
and $q_{nn'}(f_1\otimes f_2)=0$ unless $n=-n'$, in which case
Eq.~(\ref{eq:zeroco}) holds. Furthermore, these conclusions hold for
arbitrary $f_i\in\DD(\OO_i)$, and we may therefore deduce that the only
$\gamma_{nn'}^{\epsilon\epsilon'}$ which can fail to vanish are those
with $n=-n'$, $\epsilon=-\epsilon'$ (which contribute to the zero
frequency part of $F_0(\tau)$). As noted at the beginning of this
section, this implies that $\Gamma$ is translationally invariant. The
remaining condition Eq.~(\ref{eq:zeroco}) simply expresses the fact
that $\Gamma$ must vanish near the diagonal set $S$. $\square$

\sect{Conclusion}

In this paper we have discussed an apparently new unique continuation
phenomenon for weak Klein--Gordon bisolutions on a cylinder: a
bisolution vanishing in a neighbourhood of a sufficiently large diagonal
set must be globally translationally invariant. It
would be interesting to see how far Theorem~\ref{Thm:main} can be
generalised. There seem to be no essential difficulties 
in extending to the Klein--Gordon equation for
more general constant Lorentzian metrics on $M$, subject to an
appropriate modification of the diagonal sets $S_{pp'}$. For 
nonconstant 
metrics there are no nontrivial translationally invariant bisolutions to
the corresponding Klein--Gordon equation, and we speculate that a strict
unique continuation result might hold in this case. It may be
that an approach based on Carleman inequalities (see e.g., Chapter~17
in~\cite{Horm}) would be fruitful here. 
Finally, one may prove analogues of our result in 2-dimensional
Minkowski space~\cite{FHK} using more standard Fourier methods. Although
a straightforward generalisation to higher dimensions is not
possible~\cite{FHK}, it would be interesting to determine whether there
are nonetheless analogues of this result. 

{\noindent\em Acknowledgments:} The need for Theorem~\ref{Thm:main}
arose in joint work with Atsushi Higuchi and Bernard Kay~\cite{F,FHK}
and I am grateful to them for discussions on this subject. I am also
grateful to Maurice Dodson for introducing me to the work of Beurling
employed here. This work was supported by EPSRC Grant
No.~GR/K~29937 to the University of York.


\begin{thebibliography}{zz}

\bibitem{Horm}   L. H\"{o}rmander,
                 {\em The analysis of linear partial differential
                 operators, Vols I--IV},
                 (Springer, Berlin, 1990--94)


\bibitem{F}      C.J. Fewster,
                 ``Bisolutions to the Klein--Gordon equation and quantum
                 field theory on 2-dimensional cylinder spacetimes'',
                 {\tt gr-qc/9804012}

\bibitem{FHK}    C.J. Fewster, A. Higuchi and B.S. Kay,
                 ``How generic is F-locality? Examples and counterexamples''
                 In preparation

\bibitem{Wald}   R.M. Wald,
                 {\em Quantum field theory in curved spacetime and black
                 hole thermodynamics}, 
                 (University of Chicago Press, Chicago, 1994)

\bibitem{Beu}    A. Beurling, 
                 ``Mittlag-Leffler Lectures on harmonic analysis (1977-1978)''
                 in {\em The collected works of Arne Beurling, Volume 2:
                 Harmonic analysis}, 
                 L. Carleson {\em et al.} (eds) 
                 (Birkh\"auser, Boston, 1989)
                 
\bibitem{Landau} H.J. Landau,
                 ``Sampling, data transmission and the Nyquist rate''
                 Proc. IEEE {\bf 55} (1967) 1701--1706
\end{thebibliography}
\end{document}